\documentclass[twoside,12pt]{article}
\usepackage{epsfig}

\newcommand{\be}{\begin{equation}}
\newcommand{\ee}{\end{equation}}
\newcommand{\bea}{\begin{eqnarray}}
\newcommand{\eea}{\end{eqnarray}}
\newcommand{\nn}{\nonumber}
\newcommand{\dmu}{\partial_\mu}

\newcommand{\lc}{{\cal L}}

\newcommand{\sqd}{\sqrt{2}}

\begin{document}

\title{ \vspace{1cm} Meson Resonances in the Open and Hidden Charm Sectors}
\author{D.~Gamermann$^1$, E.~Oset$^1$, D.~Strottman$^1$ and M. J. Vicente Vacas$^1$
\\
{\small{\it $^1$Departamento de F\'{\i}sica Te\'orica and IFIC,
Centro Mixto Universidad de Valencia-CSIC,}}\\
{\small{\it Institutos de
Investigaci\'on de Paterna, Aptdo. 22085, 46071 Valencia, Spain}}
}

\maketitle
\begin{abstract} 
We briefly expose our model for generating open and hidden charm resonances and present the most interesting results.
\end{abstract}

\section{Introduction}

The interpretation of mesons as $q\bar q$ is already known to be insufficient in order to describe the whole meson spectrum. In particular, many of the new discovered resonances in the open and hidden charm sectors \cite{exp1,exp2} are not well described by quark models. This situation has opened the discussion about the structure of such states. 

The use of chiral Lagrangians in unitarized coupled channels has already been proved efficient in order to describe some of the states where the simple $q\bar q$ picture fails. In these cases the states are generated dynamically, through the interaction and appear as poles in the complex scattering T-matrix. 

In this work we present results coming from the use of phenomenological Lagrangians \cite{meu1,meu2} describing the interaction of pseudoscalar and vector mesons in coupled channels. The Lagrangians are based in the $SU(4)$ flavor symmetry but, since $SU(4)$ is not a very good symmetry of nature, it will be explicitly broken by taking into account the heavy mass of the exchanged mesons in the underling dynamics of the interaction. In next section the Lagrangians used are presented and the mathematical framework is briefly explained. Section 3 presents the most important results.


\section{Framework}

The pseudoscalar and the vector mesons are represented by a 15-plet representation of $SU(4)$. First two fields are constructed by means of the $SU(4)$ generators:

\be
\Phi=\sum_{i=1}^{15}{\varphi_i \over \sqd}\lambda_i;\textrm{ }{\cal{V}}_\mu=\sum_{i=1}^{15}{v_{\mu i} \over \sqd}\lambda_i \nn
\ee

Each one of these fields is a 4x4 matrix. The meson assignment for each element of the matrices can be found in \cite{meu1,meu2}. Now for each field we define a hadronic current:

\be
J_\mu=(\dmu \Phi)\Phi-\Phi\dmu\Phi;\textrm{ }\cal{J}_\mu=(\dmu \cal{V}_\nu)\cal{V}^\nu-\cal{V}_\nu\dmu \cal{V}^\nu . \label{curj}
\ee

The Lagrangians are built by connecting these currents:

\be
\lc_{PPPP}={1\over12f^2}Tr({J}_\mu {J}^\mu+\Phi^4 M);\textrm{ }\lc_{PPVV}={-1\over 4f^2}Tr\left(J_\mu\cal{J}^\mu\right). \label{lag}
\ee

In the first Lagrangian of eq. (\ref{lag}) we take $M=\textrm{diag}\left(m_\pi^2,m_\pi^2,2m_K^2-m_\pi^2,2m_D^2-m_\pi^2\right)$, in this way we reproduce the lowest order chiral Lagrangian in the light sector.

In order to break $SU(4)$ symmetry, processes driven by heavy mesons are suppressed by the factors $\gamma=\left(m_L\over m_H\right)^2$ or $\psi=-{1\over3}+{4\over3}\left(m_L\over m'_H\right)^2$. Where $m_L$, $m_H$ and $m'_H$ are typical masses of light, charmed and hidden charm vector mesons, respectively.

From the Lagrangians in (\ref{lag}) one get tree level transition amplitudes between any possible initial and final state. These amplitudes are projected in s-wave and used in the Bethe-Saltpeter equation which in an on-shell formalism is an algebraic equation:

\be
T=(\hat 1-VG)^{-1}V.
\ee
where $G$ is the loop function and $V$ a matrix containing the transition amplitudes for the different processes in all possible coupled channels.

Solving this equation in the complex plane, poles are generated in some sectors and these poles can be associated with resonances.


\section{results}

Our model generates dynamically poles which can be associated to most of the known scalar and axial resonances. In particular the $D_{s0}^*(2317)$, $D_{s1}(2460)$ and $X(3872)$ are well reproduced. Interesting predictions of our model are a heavy hidden charm scalar with mass around the $D\bar D$ threshold and an axial with negative G-parity state almost degenerated in mass with the $X(3872)$. 

Also broad open charm states with exotic quantum numbers appear. These results are, though, less stable and contrast with previous works done where states with the same quantum numbers are narrower \cite{kolo, hoff, guo1, guo2}. This sector seems to be model dependent and such predictions should be taken with care.

\end{document}